\documentstyle[floats,prd,aps]{revtex}
\begin{document}
\draft
\preprint{CWRU-P14-95}
\title{Time and chaos in general relativity }
\author{ Neil J. Cornish}
\address{Department of Physics, Case Western Reserve University, Cleveland,
Ohio 44106-7079}
\twocolumn[
\maketitle
\widetext
\begin{abstract}
The study of dynamics in general relativity has been hampered by a lack of
coordinate independent measures of chaos. Here we present a variety of
invariant measures for quantifying chaotic dynamics in relativity by
exploiting the coordinate independence of fractal dimensions. We discuss
how preferred choices of time naturally arise in chaotic systems and how
the existence of invariant signals of chaos allow us to reinstate
standard coordinate dependent measures. As an application, we study
the Mixmaster universes and find it to exhibit transient soft chaos.
\end{abstract}
]
\narrowtext

The last century has seen the emergence of three branches of physics
which have combined to overthrow the Newtonian ideal of a clockwork
universe. Our new world view encompasses an uneasy alliance of quantum
mechanics, relativity and, to a lesser extent, chaos theory. One of the most
remarkable aspects of these theories, in their current form, is their marked
mutual incompatibility. The existence of a minimum area in the phase space
of quantum mechanics, $\hbar$, suppresses chaos, while the
ubiquity of chaos in classical systems demands a new description of
the semi-classical regime. The privileged role played by time in both
quantum mechanics and chaos theory has led to a head-on collision with
general relativity where the choice of a time coordinate is arbitrary.

In this letter we offer at least a partial solution to one of these
problems, namely, the classification and quantification of chaos in
general relativity. Extensive discussion of the problem can be found in
Ref.\cite{burd}, so we shall limit our discussion to a brief review.
In general relativity both space and time are dynamical and intermixed.
There is no such thing as {\em the} time direction. In contrast, chaos
theory has been developed for Newtonian dynamics where time and space
are absolute and the notion of a mechanical phase space is clear.
It may appear that the standard tools of chaos theory can be directly
applied to relativity in schemes such as the ADM formalism\cite{adm} where
an explicit space-time split is made. This is not the case. The
coordinate freedom remains in the choice of lapse and shift functions\cite{adm}
which describe the $3+1$ decomposition. The fundamentally different
role played by time in relativity and Newtonian mechanics manifests
itself in the coordinate, or gauge, dependence of chaotic
measures such as Lyapunov exponents\cite{os}. Lyapunov exponents
are the standard method for quantifying chaotic behaviour as
they directly measure sensitive dependence on initial conditions. If
two initially close trajectories separate along a given eigendirection as
\begin{equation}
\varepsilon(t)=\varepsilon_{0} e^{\lambda t} \; ,
\end{equation}
then $\lambda$ represents the Lyapunov exponent along that direction.
If $\lambda>0$ the system is said to exhibit sensitive dependence on
initial conditions with a characteristic chaotic, or Lyapunov, timescale
$T_{L}=1/\lambda$. Unfortunately, this nice picture breaks down when
applied to general relativity. Consider the allowed coordinate transformation
$t \rightarrow \ln \tau $. In terms of this time variable we find
\begin{equation}
\varepsilon(\tau)=\varepsilon_{0} \tau^{\lambda} \; ,
\end{equation}
which describes the standard power-law divergence of trajectories found in
integrable system. In particular, the Lyapunov exponents in
this coordinate system would all be zero. It should be mentioned
that the Lyapunov exponents also depend on the choice of distance measure
in phase space and are therefore variant under spatial coordinate
transformations also. From the above discussion it is clear that standard
coordinate dependent measures of chaos have to be either modified,
abandoned or augmented in general relativity\cite{rugh}.

In order to find methods suitable for classifying and quantifying chaos
in general relativity we shall exploit the remarkable connection between
chaos and fractal curves. The utility of these methods comes from the
coordinate or diffeomorphism invariance of fractal dimensions. Fractal
structures can be found in the phase space of all chaotic system. They
may be uncovered by taking Poincar\'{e} sections, plotting attractor basin
boundaries or by finding the intersections of stable and unstable phase space
manifolds. These methods reveal respectively, cantori\cite{aub}, fractal basin
boundaries\cite{greb} and chaotic invariant sets\cite{kan}. Actually, fractal
basin boundaries are a particular type of chaotic invariant set. The
connection between chaos and fractals is deep. A non-chaotic, integrable
system has sufficient isolating integrals (constants of the motion) to
fully determine the dynamics. The trajectories of an integrable system are
restricted by these isolating integrals to lie on smooth manifolds in
phase space with the topology of $n$-dimensional tori.
In chaotic systems there are insufficient isolating integrals and the
smooth tori are replaced by fractal cantori - locally the product of a torus
and a Cantor set. The fractal dimension of a cantorus captures topological
information\cite{foot} about a trajectory while Lyapunov exponents
measure metrical properties. The importance of such topological information
as a qualitative sign of chaos in general relativity has been emphasised
by Calzetta and El Hasi\cite{cal} and Dettmann {\it et. al.}\cite{carl}. In
what follows, we discuss quantitative relationships that relate
fractal dimensions to important quantities such as final state sensitivity,
Lyapunov exponents and chaotic entropy.

A unified description of chaotic dynamics is possible in terms of
chaotic invariant sets\cite{ott}. Chaotic invariant sets are formed by
the intersection of stable and unstable manifolds in phase space. The
complex, and often fractal, nature of these sets follows from the result
that if a stable and unstable manifold intersect once, they intersect and
infinite number of times. Familiar examples of chaotic invariant sets are
strange attractors and fractal basin boundaries. A lesser known example
is the strange repeller responsible for chaotic scattering\cite{bl,ll}.

For low-dimensional systems an intriguing result has been
found relating the fractal dimension of chaotic invariant sets to the their
Kolmogorov-Sinai entropy\cite{ks} and Lyapunov exponents. The importance of
these relations in general relativity follows from the diffeomorphism
invariance of the multifractal dimensions $D_{q}$. The fractal
dimensions are defined by
\begin{equation}\label{dimdef}
D_{q}={1 \over q-1} \lim_{\epsilon \rightarrow 0} {\ln 
\sum_{i=1}^{N(\epsilon)} (p_{i})^{q} \over \ln \epsilon } \; ,
\end{equation}
where $N(\epsilon)$ are the number of hypercubes of side length $\epsilon$
needed to cover the fractal and $p_{i}$ is the weight assigned to the
$i^{{\rm th}}$ hypercube. The $p_{i}$'s satisfy
$\sum_{i=1}^{N(\epsilon)} p_{i} = 1$.
The standard capacity dimension is
recovered when $q=0$, the information dimension when $q=1$, the correlation
dimension when $q=2$ etc. For homogeneous fractals all the various
dimension yield the same result. The multifractal dimensions $D_{q}$ are
invariant under diffeomorphisms for all $q$, and $D_{1}$ is additionally
invariant under coordinate transformations that are non-invertible at
a finite number of points\cite{owy}. If a strange attractor or repeller has
a multifractal spectrum $D_{q}$ with continuous first derivative
$\partial_{q}D_{q}$ it is said to be uniformly hyperbolic. Physically
this means that all trajectories on the attractor or repeller are
unstable. In practice, most attractors and repellers are not
uniformly hyperbolic and exhibit a phase transition at some
$q=q_{{\scriptscriptstyle T}}$ where
$D_{q}$ has a discontinuous derivative. The breakdown of hyperbolicity 
for $q>q_{{\scriptscriptstyle T}}$ is due to the presence of stable orbits
in an otherwise chaotic system. Such systems are said to exhibit soft chaos.

Remarkably, it has been shown that $D_{1}$ equals the Lyapunov dimension
$D_{L}$ for 2-dimensional systems.
The Lyapunov dimension is defined by
\begin{equation} \label{main}
D_{L}=h(\mu)\left( {1 \over \lambda_{1}}-{1 \over \lambda_{2}}\right) \; ,
\end{equation}
where $\lambda_{1}>0>\lambda_{2}$ are the Lyapunov exponents in each
eigendirection and $h(\mu)$ is the metric or K-S entropy of the chaotic
invariant set with respect to the set's natural measure $\mu$
(see Ref. \cite{ott} or Ref. \cite{ll} for definitions of $h(\mu)$,
$\lambda$ and $\mu$). The metric entropy is bounded from above by the
topological entropy $H$\cite{bar}. As its name suggests,
the topological entropy is diffeomorphism invariant while the metric
entropy can be transformed to zero by a coordinate change.
In practice, the topological entropy is much harder to calculate
than the metric entropy, but when it can be calculated, it provides
a gauge invariant signal of chaos in addition to the $D_{q}$'s.

The relation $D_{1}=D_{L}$ has been rigorously
established for certain dynamical systems\cite{young} and has been
numerically confirmed for many typical systems\cite{kan}. Similar relations
have been conjectured to hold for $n$-dimensional systems\cite{ky}. In
Hamiltonian systems conservation of phase space volume implies
$\lambda_{1}=-\lambda_{2}$. For typical strange attractors and repellers the
metric entropy is given by
\begin{equation}
h(\mu)=\lambda_{1}-{1 \over \tau_{d}} \; ,
\end{equation}
where $\tau_{d}$ is the decay time for trajectories leaving the repeller.
Strange repellers are found in systems displaying chaotic
scattering, examples of which have been explored in general
relativity\cite{carl,conto,cos}. For dissipative systems with strange
attractors $\tau_{d}=\infty$ and (\ref{main}) reduces to the Kaplan-Yorke
relation\cite{ky}. The Lyapunov dimension of a strange repeller is
a particularly useful measure of chaos as it
does not require a compact phase space. Most dynamical systems in general
relativity do not have compact phase spaces.

With these preliminaries out of the way, we can now focus on the important
implications eqn.(\ref{main}) has for chaos in general relativity. The result
is direct: although neither Lyapunov exponents nor K-S entropies are
coordinate invariant, their ratio is. This result allows us to reinstate
both Lyapunov exponents and K-S entropies as useful chaotic measures
in a fixed gauge, so long as we first verify that $D_{1}$ has a
fractional value. In practice, it is easy to reconstruct
chaotic invariant sets numerically\cite{pim} and measure their fractal
dimension. We use the fractal value of $D_{1}$ to prove the system
is chaotic. Of course, even for chaotic systems there are an infinite
number of gauge choices in which the Lyapunov exponents and K-S entropy
vanish. Therefore, as a physical choice, we exclude all
coordinate choices where chaotic systems have $D_{1}=D_{L}=0/0$. On
reflection, it is clear that a preferred time choice is a natural consequence
of chaos as the K-S entropy selects a chaotic arrow of time.
In the confines of a good gauge choice, we can compare quantities such as
Lyapunov times to the
characteristic timescales of other physical processes. In practice these are
the physically important, albeit gauge variant, questions we need to ask.

A related method of choosing non-pathological gauges employs Poincar\'{e}
sections. Regardless of the choice of gauge, chaotic trajectories will
appear as fractal curves (cantori and stochastic layers) while stable
trajectories will appear as smooth curves (KAM tori). Pathological gauge
choices can
be defined as those which have vanishing Lyapunov exponents for cantori
or non-vanishing exponents for KAM tori. At present there appears to
be no way to quantitatively relate the fractal dimension of cantori to
their Lyapunov exponents. Some preliminary results have recently been found
relating the gap structure of cantori to their Lyapunov exponents\cite{co},
and hopefully new results will soon emerge to bolster our qualitative picture.

For those uncomfortable with preferred time choices, there is another
method for quantifying chaos that is entirely gauge independent.
A defining feature of chaotic systems is their sensitive dependence
on initial conditions. Generally, a range of possible outcomes can be
assigned to a dynamical system, and each outcome has a basin of
attraction in the space of initial conditions. For chaotic systems the
basin boundaries are fractal, and the fractal dimension of the boundary
provides a coordinate independent measure of the chaotic dynamics
\cite{carl,cos}. The quantitative importance of the fractal dimension
is expressed in terms of the final state sensitivity $f(\delta)$\cite{mc}.
This quantity describes how the unavoidable uncertainty in
specifying initial conditions gets amplified in
chaotic systems, leading to a large final state uncertainty. The function
$f(\delta)$ is the fraction of phase space volume which has an
uncertain outcome due to the initial conditions being uncertain within
a hypersphere of radius $\delta$. It can be shown\cite{mc} that
\begin{equation}\label{final}
f(\delta)\sim \delta^{\alpha}\;, \quad \alpha=N-D_{0} \; ,
\end{equation}
where $N$ is the phase space dimension and $D_{0}$ is the capacity dimension
of the basin boundary. For non-chaotic systems $\alpha=1$ and there is no
amplification of initial uncertainties, while for chaotic systems
$0< \alpha <1$ and marked final state sensitivities can occur. For example,
if $\alpha=0.1$ a $50 \% $ reduction in the initial uncertainty only results
in a $7 \% $ reduction in the final state uncertainty. By reversing the
argument, eqn.(\ref{final}) can be used to quickly determine the dimension
of the basin boundary\cite{mc}.

One application for the gauge invariant measures $D_{L}$, $H$, and $f(\delta)$
advanced in this letter might be to settle the long running debate over
the existence of chaos in the Mixmaster universe\cite{burd,bar}. At most,
the Mixmaster will only exhibit transient chaos as its trajectories are
asymptoically regular\cite{mw}. To show the Mixmaster is chaotic, we need
only consider the reduction of the full dynamics to the discrete $(u,v)$ map,
as this map accurately describes the majority of Mixmaster
trajectories\cite{bb}. The map is defined by
\begin{equation}
(u_{n+1},v_{n+1})=\left\{ \begin{array}{ll}
(u_{n}-1, v_{n}+1) & u_{n}>2\, , \\ & \\
{\displaystyle \left({1 \over  u_{n}-1} , 
{v_{n} \over v_{n}+1}\right)} & 1< u_{n} < 2\, .
\end{array} \right.
\end{equation}
Using the standard metrical measures of chaos it is unclear whether or
not the $(u,v)$ map is chaotic. For a finite number of iterations, $n$,
the map has a short-time Lyapanov exponent given by
$\lambda_{n}=\pi^2 / (6 \ln 2 \ln u_{{\rm max}})$,
where $u_{\max}$ is the largest value of $u$ visited during the $n$
itera-
\newpage

\
\begin{figure}[h]
\vspace{50mm}
\includegraphics{blank.ps}
\includegraphics{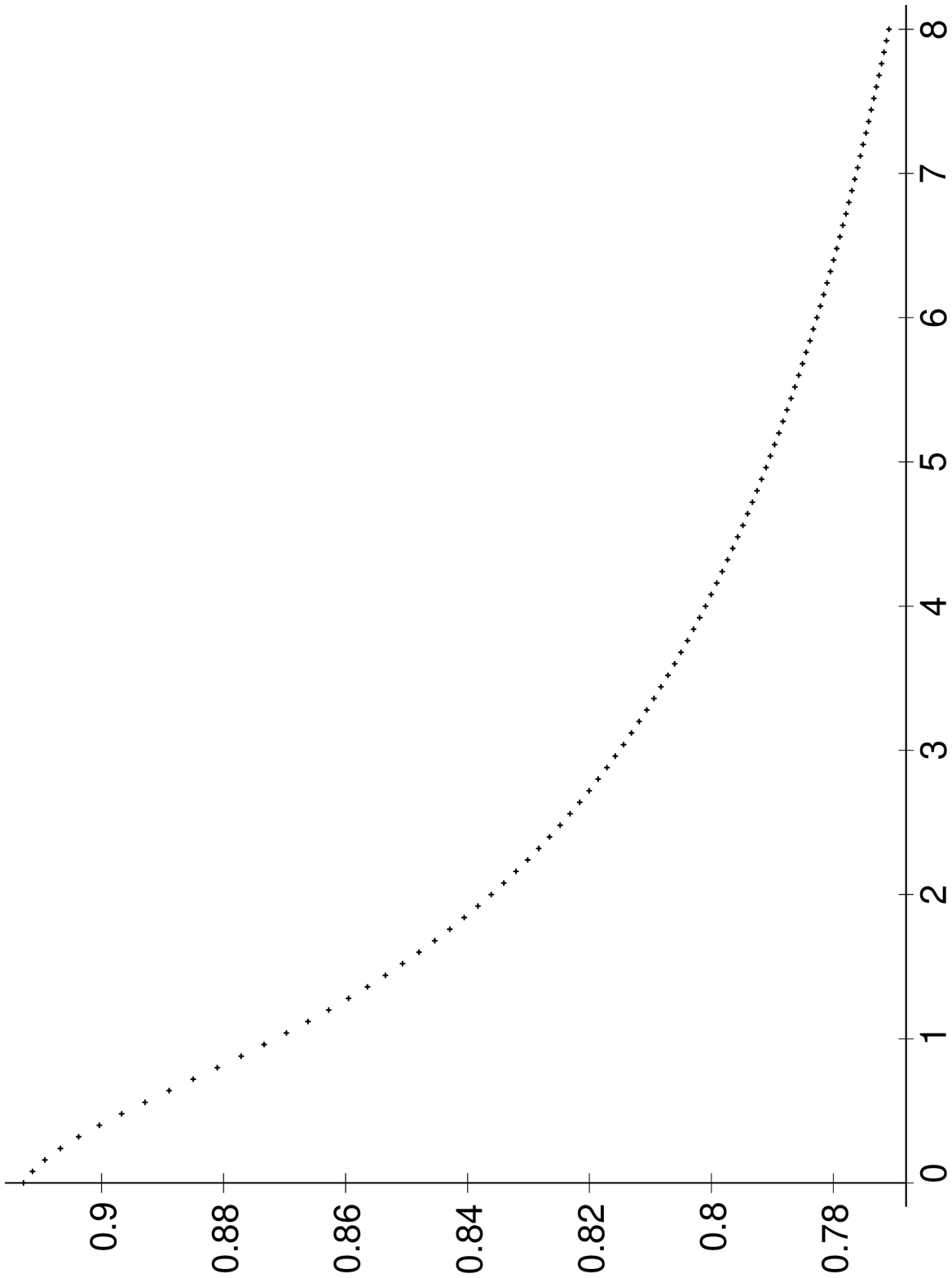}
\vspace{8mm}
\caption{The multifractal dimension $D_{q}$ as a function of $q$.}
\end{figure} 

\begin{picture}(0,0)
\put(-6,200){$D_{q}$}
\put(125,45){$q$}
\end{picture}

\noindent 
tions\cite{bb}. However, as we take the limit $n\rightarrow \infty$ to
recover the true Lyapanov exponent we find for typical trajectories that
$u_{{\rm max}} \rightarrow \infty$ and $\lambda=0$. This regular
asymptotic behaviour, combined with positive short-time Lyapanov
exponents is the hallmark of chaotic scattering. For the $(u,v)$ map
the chaotic scattering occurs when $1<u<2$. In order to prove that
the $(u,v)$ map harbours a strange repeller we must reconstruct its
chaotic invariant set. For a map this corresponds to the set of fixed points
$(u_{n+k},v_{n+k})=(u_{n},v_{n})$ for all integers $k$. By numerically
generating this set we are able to find its spectrum of multifractal
dimensions $D_{q}$. Since the $(u,v)$ map
is invertible, we need only consider the chaotic future invariant set
generated by the $u$ map. The topological entropy of the set is given by
\begin{eqnarray}
H&=&\lim_{k\rightarrow \infty} {1 \over k} \ln \left[ {\rm No.\; of\; fixed\;
points\; at\;  order\; } k\right] \nonumber \\ 
&=& \ln 2    \, .
\end{eqnarray}
The result $H=\ln 2$ follows from a simple counting argument.
Since $H > 0$ we are assured that the $(u,v)$ map is chaotic and
harbours a strange repeller. It is interesting to note that the Mixmaster
$(u,v)$ map has the same topological entropy as the Smale Horseshoe.
As we would expect, the compactified version of the $(u,v)$ map,
the Gauss map, has a considerably larger topological entropy of
$H_{G}=\pi^2 /6/(\ln 2)^2$\cite{bar}.
Numerically the topological entropy was
found to converge to $\ln 2$ very quickly with $k$, differing by less than
1 part in 1000 for $k\geq 10$. To be on the safe side, we chose the finite
approximation $k=\{1..15\}$ when calculating the multifractal dimensions. The
task of calculating $D_{q}$ is made easier by the dense nature of
the repeller for small $u$. We find the fraction of all fixed points in
a given integer interval $[1,u]$ to be
\begin{equation}
F(u)=1-2^{-u+1} \; ,
\end{equation}
so the core of the strange repeller is strongly localised around $u=1$.
For this reason, we chose to measure the fractal dimension in the
interval $u=[1..10]$ as it contains $99.8\%$ of the repeller. The
above approximations are particularly good for evaluating multifractal
dimensions with positive $q$ as these focus on the dense regions of
a fractal. Conversely, our truncation to finite $k$ and $u$ will
lead to large errors as $q\rightarrow -\infty$.
The spectrum of dimensions $D_{q}$ for the $u$ map is displayed
graphically in Fig. 1.
For reference, $D_{0}=0.91\pm 0.01$ and $D_{1}=0.872\pm 0.005$. If the map
is hyperbolic at $q=1$ we would expect to find
$D_{1}=1-1 /( \lambda\, \tau_{d})$,
in accordance with eqn.({\ref{main}). However, the decay of orbits from
the repeller proceeds as
\begin{equation}
N(n)=N_{1}\exp(-n/\tau_{d})+N_{2}(n+1)^{-a} \; .
\end{equation}
The power-law tail is typical for repellers punctured by stable periodic
orbits, and leads to what is known as soft chaos. This indicates that
the repeller is not hyperbolic at $q=1$ and we cannot expect eqn.(\ref{main})
to hold. Since the repeller is not uniformly hyperbolic, it would be
interesting to verify that the $D_{q}$ undergo a phase transition.
Unfortunately, the enhanced numerical uncertainties for $q<1$
makes this very difficult to study. While it is disappointing that the
Mixmaster is one of the rare systems where eqn.(\ref{main}) does not hold, we
have nonetheless been able to extract useful information
about the dynamics by measuring the multifractal dimensions
and topological entropy of the underlying strange repeller.
To summarise, we have shown that the Mixmaster universe exhibits
transient soft chaos due to a weak strange repeller. Our fractal
approach has suceeded where metrical methods have failed.

On a grander
scale, it may be that chaos theory has a role to play in reconciling quantum
mechanics and general relativity in the context of quantum cosmology\cite{hal},
where chaos related phenomena such as decoherence, Fokker-Planck diffusion
and dynamical arrows of time are thought to be important.

This work builds on my collaborations with Carl Dettmann, Sam Drake,
Norm Frankel and Janna Levin. I would like to thank Albert Fathi, Robert
MacKay and Edward Ott for answering several questions concerning cantori
and dimensions.


\begin{references}
\bibitem{foot} The fractal dimension is not a true topological invariant
as it is only invariant under diffeomorphisms, not homeomorphisms. This
distinction will not concern us as we only need to be concerned about
diffeomorphisms in general relativity.
\end{references}
\end{document}